# Charge density wave origin of cuprate checkerboard visualized by scanning tunneling microscopy

W. D. Wise[1], M. C. Boyer[1], Kamalesh Chatterjee[1], Takeshi Kondo[2,1,*], T. Takeuchi[2,3], H. Ikuta[2], Yayu Wang[1,†], E. W. Hudson[1]

[1]Department of Physics, Massachusetts Institute of Technology, Cambridge, MA 02139, USA.
[2]Department of Crystalline Materials Science, Nagoya University, Nagoya 464-8603, Japan.
[3]EcoTopia Science Institute, Nagoya University, Nagoya 464-8603, Japan.
*Present address: Ames Laboratory and Dept. of Physics and Astronomy, Iowa State University, Ames, IA 50011, USA.
†Present address: Department of Physics, Tsinghua University, Beijing 100084, China.

**One of the main challenges in understanding high $T_C$ superconductivity is to disentangle the rich variety of states of matter that may coexist, cooperate, or compete with $d$-wave superconductivity. At center stage is the pseudogap phase, which occupies a large portion of the cuprate phase diagram surrounding the superconducting dome[1]. Using scanning tunneling microscopy, we find that a static, non-dispersive, "checkerboard"-like electronic modulation exists in a broad regime of the cuprate phase diagram and exhibits strong doping dependence. The continuous increase of checkerboard periodicity with hole density strongly suggests that the checkerboard originates from charge density wave formation in the anti-nodal region of the cuprate Fermi surface. These results reveal a coherent picture for static electronic orderings in the cuprates and shed important new light on the nature of the pseudogap phase.**

A great deal of current interest is focused on the "checkerboard"-like electronic lattices first discovered in cuprates by scanning tunneling microscopy (STM) in vortex cores in optimally doped $Bi_2Sr_2CaCu_2O_{8+\delta}$ (Bi-2212)[2]. This ordering was found to have a roughly four unit-cell ($4a_0$) wavelength orientated along the Cu-O bond direction. Subsequent STM investigations of the cuprates have revealed other checkerboard structures in the absence of a magnetic field. For example, in the superconducting state of Bi-2212 the first report of a checkerboard saw a roughly $4a_0$ wavelength throughout the sample[3], while a later study found the ordering (wavelength $4.5a_0$) limited to regions with large gap ("zero-temperature pseudogap") tunneling spectra[4]. A checkerboard was also found in slightly underdoped Bi-2212 above the superconducting transition temperature $T_C$ with wavelength $4.7a_0 \pm 0.2a_0$ (Ref. 5). In $Ca_{2-x}Na_xCuO_2Cl_2$ (Na-CCOC), a commensurate electronic crystal phase with period $4a_0$ was found at low temperatures in both superconducting and non-superconducting samples[6].

Although it is not yet clear whether these checkerboards all represent the same electronic entities, many models have been proposed to explain the mechanisms of these novel electronic phases and their implications for the pseudogap and high $T_C$ superconductivity[7-17]. Initially, it was suggested[2,9] that the $4a_0$ pattern in Bi-2212 vortex cores is the charge density modulation accompanying the $8a_0$ spin density wave (SDW) created by an external magnetic field. Other explanations of checkerboards include exotic orderings such as fluctuating one-dimensional stripes[12], modulations of electron hopping amplitude[13], Wigner crystal or charge density wave of Cooper pairs[14,15], and orbital current induced $d$-density wave[17]. Recently, angle resolved photoemission spectroscopy (ARPES) on Na-CCOC found parallel Fermi surface (FS) segments with nesting vector around $2\pi/4a_0$ in the anti-nodal region, suggesting charge density wave (CDW) formation as the origin of the checkerboard[18]. Unfortunately, existing data are inadequate to discriminate between the different models, mainly because the experiments were carried out on small, isolated regions of the complex cuprate phase diagram.

Here we report on systematic doping- and temperature-dependent STM studies of charge density modulations in the high temperature superconductor $Bi_{2-y}Pb_ySr_{2-z}La_zCuO_{6+x}$ (Pb and La substituted Bi-2201)[19]. We find that a static (non-fluctuating), non-dispersive (energy independent), checkerboard-like electronic lattice exists over a wide range of doping, and that its wavelength increases with increasing hole density. This unexpected trend strongly supports the physical picture of Fermi surface nesting induced charge density wave formation and is corroborated by comparison to band structure calculations and ARPES measurements.

These experiments were conducted on a home-built variable temperature STM, which allows simultaneous mapping of atomic scale topography and differential conductance spectroscopy, proportional to the energy dependent local density of states (LDOS) of the sample. We begin by describing our results on optimally doped Bi-2201 with $T_C$ = 35 K. Figure 1a is a typical atomic resolution STM topography of a 785 Å region measured at $T$ = 6 K. The inset shows the Pb (brighter) and Bi (dimmer) atoms of the exposed BiO plane. The $CuO_2$ plane lies ~ 5 Å below. A representative differential conductance spectrum from this area (Fig. 1b) has a clear inner gap with peaks near 15 meV, probably associated with the superconducting gap, and a pseudogap with size roughly 75 meV[20]. A differential conductance map of the region taken at a bias of 10 meV (Fig. 1c) shows a checkerboard-like electronic lattice, strikingly similar to those observed in other cuprates[4-6]. The checkerboard is observed to beyond 50 meV at both positive and negative sample bias, although the pattern appears most strongly at low, positive bias. It appears in maps taken with feedback setpoint voltages ranging from 10 mV to 300 mV, with feedback currents from 50 pA to 800 pA, and in topographic scans at 10 mV bias.

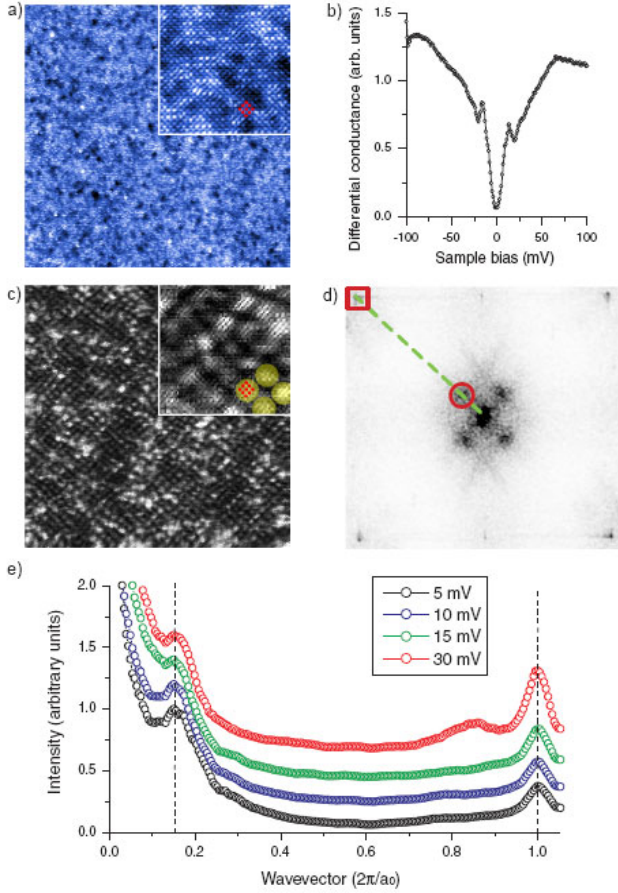

**Fig. 1. Anatomy of the checkerboard in optimally doped Bi-2201.** **(A)** STM topography (785 Å) of optimally doped ($T_C$ = 35 K) Bi-2201 measured at $T$ = 6 K. The magnified inset (110 Å) (and red dots on 9 atoms) show the clear atomic lattice in this high resolution data. **(B)** Spatially-averaged differential conductance spectra measured in the area shown in (A) exhibit two distinct gaps: a superconducting gap $\Delta_{SC}$ ~ 15 meV and pseudogap $\Delta_{PG}$ ~ 75 meV. **(C)** Conductance maps, here taken with bias voltage 10 meV on the same region as (A), show a checkerboard structure in the local density of states with a wavelength much larger than the atomic lattice. The inset is magnified as in (A), with the same 9 atoms highlighted in red. Four checkerboard maxima are also highlighted (yellow) for clarity. **(D)** Fourier transform of the map shown in (C). The checkerboard wavevectors (circled) appear as four spots along the same direction as the atomic lattice (boxed). The dashed line shows the locations of the line cuts **(E)** extracted from FT-LDOS maps with different bias voltages. The left vertical line marks the position of the checkerboard wavevector, $2\pi/6.2a_0$ for all energies, and the right vertical line indicates the atomic lattice wavevector $2\pi/a_0$. All data in this paper was acquired with feedback setpoint parameters $I_S$ = 400 pA and $V_S$ = -100 mV or $V_S$ = -200 mV.

The wavelength of this checkerboard is determined from the Fourier transform (FT) of the image, as shown in Fig. 1d, where the checkerboard appears as four peaks (one is circled). Its wavevector corresponds to a wavelength $d$ ~ $6.2a_0 \pm 0.2a_0$, much larger than that of any such structure previously reported. Fig. 1e shows a line cut along the atomic lattice $(\pi, 0)$ direction of the FT-LDOS maps taken at different bias voltages. The consistent position of the checkerboard wavevector observed at different energies, marked by the left broken line, indicates that the checkerboard is a non-dispersive, static ordering.

We find similar checkerboard structures in underdoped Bi-2201 samples with $T_C$ = 32 K (Fig. 2b) and $T_C$ = 25 K (Fig. 2c). Surprisingly, FTs reveal that checkerboard periodicities in these underdoped samples are reduced to $5.1a_0 \pm 0.2a_0$ and $4.5a_0 \pm 0.2a_0$ respectively, significantly shorter than in the optimally doped sample. This can be seen directly from the denser packing of the underdoped checkerboard (Fig. 2b and 2c) than that in the optimally doped one (Fig. 2a). Fig. 2d summarizes this doping dependence in line cuts of the FTs along the atomic lattice $(\pi, 0)$ direction. The increase of checkerboard wavevector with decreasing hole density is pronounced.

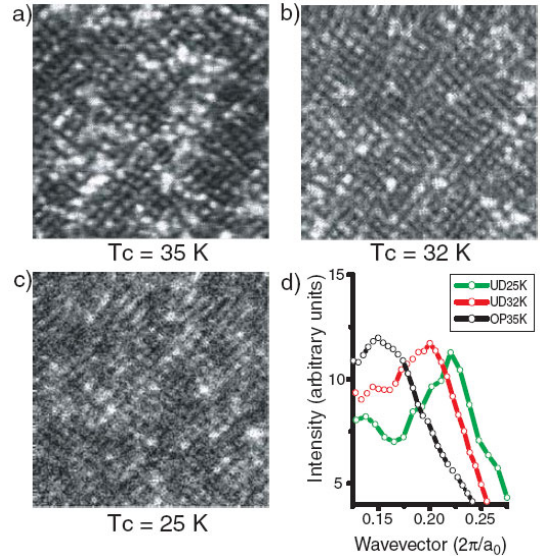

**Fig. 2. Doping dependence of the checkerboard.** 400 Å conductance maps of **(A)** optimally doped, $T_C$ = 35 K, **(B)** underdoped, $T_C$ = 32 K, and **(C)** underdoped, $T_C$ = 25 K, Bi-2201. All maps were taken with 10 mV sample bias at $T$ = 6 K. The checkerboard structures shown in (B) and (C) have denser packing than in (A), indicating a shorter wavelength in underdoped samples. **(D)** Line cuts along the atomic lattice direction of the FT-LDOS maps of the three samples. The cuts peak at the checkerboard wavevectors, corresponding to wavelengths of $6.2a_0$, $5.1a_0$, and $4.5a_0$, respectively.

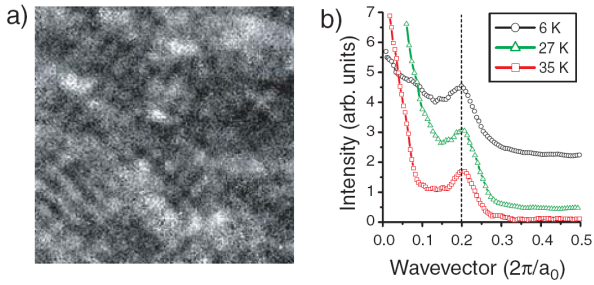

**Fig. 3. Temperature independence of the checkerboard. (A)** 300 Å, 10 mV conductance map of the underdoped $T_C$ = 32 K sample measured at $T$ = 35 K, slightly above $T_C$. The checkerboard is qualitatively unchanged from low temperatures (Fig. 2B). **(B)** Line cuts along the atomic lattice direction of FT-LDOS maps at this and lower temperatures indicate that the checkerboard wavevector is temperature independent (vertical line).

In contrast to doping, temperature has no measurable effect on the checkerboard wavevector. The LDOS map of the underdoped $T_C$ = 32 K sample measured at 35 K (Fig. 3a) is qualitatively the same as that measured at 6 K (Fig. 2a). In figure 3b, we show line cuts of the FTs of maps measured at a wide range of temperatures, demonstrating that the peak location is unaffected by temperature and in particular $T_C$.

These results reveal important new features of the checkerboard. First, the non-dispersive charge density modulation found previously in Bi-2212[4,5] and Na-CCOC[6] also exists in Bi-2201, suggesting that it is a robust feature that prevails in the cuprate phase diagram, in both optimally doped and underdoped phases, and at temperatures both below and above $T_C$. More importantly, the doping dependence of the checkerboard periodicity puts stringent constraints on relevant theoretical models, as discussed below.

We first emphasize that the checkerboard structures reported here and previously[4-6] are distinct from the spatial LDOS modulations induced by quasiparticle interference (QPI)[21]. QPI wavevectors depend strongly on energy because they are formed by interference of elastically scattered quasiparticles residing on equal-energy contours of the FS[22]. In contrast, the checkerboard lattice is non-dispersive. For example, wavevectors predicted by the standard octet model of QPI[22] would disperse from 0.18 to 0.33 (in units of $2\pi/a_0$) over the energy range shown in Fig 1, clearly inconsistent with the non-dispersing vectors we report here. Similarly, QPI from the ends of the Fermi arcs[23], a seemingly reasonable explanation of the checkerboard observed strictly above $T_C$ in Bi-2212 (Ref. 5), cannot explain the temperature independent, non-dispersive pattern reported here.

For non-dispersive checkerboard formation, a number of explanations have been proposed. Most predict a checkerboard wavelength which decreases with doping, opposite our results. For example, the presence of a $4a_0$ checkerboard in Bi-2212 vortex cores[2] was initially attributed[2,7-11] to the long-sought concomitant charge modulation of the $8a_0$ SDW found earlier by neutron scattering in the vortex lattice state of LSCO[24]. In a variety of cuprates, neutron scattering experiments have found incommensurate SDWs, which should create accompanying charge density modulations with half the wavelength[7-11]. Although this is an appealing picture for the vortex checkerboard, that the measured SDW wavelength decreases with increasing doping[25] means it cannot explain the patterns reported here.

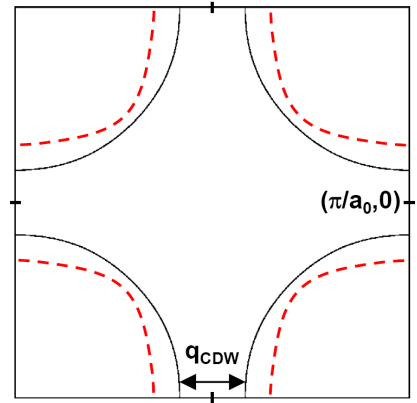

**Fig. 4. Fermi surface nesting.** Tight binding calculated Fermi surface (solid black curve) of optimally doped Bi-2201[26] based on ARPES data[27]. The nesting wavevector (black arrow) in the anti-nodal flat band region has length $2\pi/6.2a_0$. Underdoped Bi-2201 Fermi surfaces (shown schematically as red dashed line) show a reduced volume and longer nesting wavevector, consistent with a CDW origin of the doping dependent checkerboard pattern reported here.

The stripe model[12], which posits that fluctuating stripes can form a checkerboard-like pattern when pinned by impurities, also predicts that wavelengths should decrease with doping, as more holes means more stripes and a smaller average distance between them. Similarly, in theories that attribute checkerboard patterns to real space organization of Coulomb-repulsed Cooper pairs in the form of Wigner crystals[14] or Cooper pair density waves[15] one would expect that the distance between neighbouring Cooper pairs would decrease with doping as the Cooper pairs become less dilute. These predictions are also at odds with our observations.

We propose that the most likely origin for the checkerboard is the formation of an incommensurate charge density wave[16]. The cuprate FS flattens out in the anti-nodal (0, π) region and forms parallel ("nested") sections there (Fig. 4). Nesting is beneficial to the formation of CDWs as a modulation at a single wavevector can gap large sections of the FS and lower electronic energy. Tight binding calculations[26] based on ARPES measurements[27] of the FS of optimally doped Bi-2201 (solid black line), indicate a nesting wavevector of $2\pi/6.2a_0$ (arrow) in the anti-nodal region, in excellent agreement with the checkerboard periodicity of $6.2a_0$. With fewer holes in the $CuO_2$ plane in underdoped samples, the cuprate hole-like FS shrinks (broken line), leading to a larger nesting wavevector and hence smaller real space wavelength, in agreement with the doping dependence we report here.

The CDW picture can also explain the checkerboards in Bi-2212 and Na-CCOC. ARPES measurements on slightly underdoped Bi-2212 reveal an $\sim 2\pi/5a_0$ anti-nodal nesting wavevector[28], in agreement with the STM measured $4.7a_0$ checkerboard periodicity[5]. In Na-CCOC, the match of the $4a_0$ checkerboard wavelength to their ARPES measured nesting wavevector around $2\pi/4a_0$ also led researchers to the conclusion of a CDW[18]. The doping dependence sought in that study, but absent in the invariably commensurate $4a_0$ checkerboard, possibly due to lock-in of an incommensurate CDW by the crystal lattice, is revealed here in Bi-2201, where the checkerboard is incommensurate and strongly doping dependent, clearly favouring a FS nesting induced CDW picture.

If CDWs are indeed a universal feature of the cuprates, one might question why x-ray and neutron scattering searches have so far been unsuccessful in finding widespread evidence of charge modulations. This may be due to the weak, glassy nature of these modulations[29]. While x-ray has successfully detected charge modulations in $La_{2-x}Ba_xCuO_4$, the in-plane correlation length there ($\xi \sim 500$ Å)[30] is significantly longer than in Bi-2212 ($\xi \sim 90$ Å)[5], Na-CCOC ($\xi \sim 40$ Å)[6] and in the Bi-2201 samples discussed here ($\xi \sim 35$ Å), perhaps due to pinning by disorder in these samples. This difference alone would be responsible for a drop in scattering signal of a couple orders of magnitude and may thus explain the lack of corroborating results from scattering measurements.

With a CDW as the most likely source of the checkerboard, we next turn our attention to the relationship between this CDW and other physics in the system. In particular, since we observe it both above and below $T_C$, the question arises as to whether the CDW is the hidden order of the pseudogap phase. Recent discovery of a dichotomy between the nodal and anti-nodal quasiparticles, revealed by Raman and ARPES experiments[31-33], supports this conjecture. Quasiparticles near the node have a $d$-wave gap which opens at $T_C$, and hence are assumed to be responsible for $d$-wave superconductivity. Quasiparticles near the anti-node on the other hand have a large gap which is roughly temperature independent near, and exists well above, $T_C$. That this gap persists above $T_C$ demonstrates that it is the pseudogap. That this gap exists near the antinodes, where the nested FS appears responsible for the formation of the CDW we report here suggests that the pseudogap may be the CDW gap.

Objecting to this claim, some have commented that a CDW gap need not be centred on the Fermi energy and thus at least at some dopings the pseudogap should be asymmetric around it. Close observation of our spectra affirms that the pseudogap is rarely symmetric about the Fermi energy. Fig. 1b shows peaks at -88 meV and +66 meV, very asymmetric particularly considering the clear symmetry of the inner (superconducting gap) peaks at ±15 meV. This asymmetry is ubiquitous in large gap regions (see for example Fig. 1a of Ref. [5] and Fig. 2b of Ref. 4), where the pseudogap is clearly distinguishable from the symmetric superconducting gap, blunting this objection to the picture of pseudogap as CDW gap.

Despite the similarities between CDWs and features of the pseudogap, much work remains to be done before confirming or refuting this picture. The pseudogap is a rich phase, exhibiting a wide variety of phenomena and, to date, no theory has consistently explained all the results of the large number of experimental probes of its nature. That the charge density wave discussed here explains some of them is a beginning.

**Acknowledgements**
We thank J.E. Hoffman, K.M. Lang, P.A. Lee, Y.S. Lee, S. Sachdev, T. Senthil, and X.-G. Wen for helpful comments. We thank C. Lindh for coding and data analysis. This research was supported in part by a Cottrell Scholarship awarded by the Research Corporation and by the MRSEC and CAREER programs of the NSF.

**Author Contributions**
WDW, MCB and KC shared equal responsibility for all aspects of this project from instrument construction through data collection and analysis. TK grew the samples and helped refine the STM. TT and HI contributed to sample growth. YW contributed to analysis and writing of the manuscript. EWH advised.

*Correspondence and requests for materials should be addressed to E.W. Hudson, ehudson@mit.edu*